\documentclass[prd,twocolumn,nofootinbib,eqsecnum,showpacs]{revtex4}

\def\w{\Omega}

\def\be{\nopagebreak[3]\begin{equation}}
        \def\ee{\end{equation}}
        \def\ba{\nopagebreak[3]\begin{eqnarray}}
        \def\ea{\end{eqnarray}}

\def\d{{\rm d}}

\newcommand{\teta}{\rlap{\lower2ex\hbox{$\,\tilde{}$}}\eta{}}

\begin{document}

\preprint{\vbox{\baselineskip=12pt \rightline{ICN-UNAM-03/12}
\rightline{hep-th/0311089} }}

\title{Note on Self-Duality and the Kodama State}
\author{Alejandro Corichi}\email{corichi@nuclecu.unam.mx}
\affiliation{Instituto de Ciencias Nucleares\\
Universidad Nacional Aut\'onoma de M\'exico\\
A. Postal 70-543, M\'exico D.F. 04510, MEXICO \\
}
\author{Jer\'onimo Cortez}\email{cortez@nuclecu.unam.mx}
\affiliation{Instituto de Ciencias Nucleares\\
Universidad Nacional Aut\'onoma de M\'exico\\
A. Postal 70-543, M\'exico D.F. 04510, MEXICO \\
}

\begin{abstract}
An interesting interplay between self-duality, the Kodama
(Chern-Simons) state and knot invariants is shown to emerge in the
quantum theory of an Abelian gauge theory. More precisely, when a
self-dual representation of the CCR is chosen, the corresponding
vacuum in the Schr\"odinger representation is precisely given by
the Kodama state. Several consequences of this construction are
explored.
\end{abstract}

 \maketitle

\section{Introduction}
\label{sec:1}

In a quantum gauge theory in $3+1$ dimensions, there is a natural
background independent state that can be defined in the
Schr\"odinger picture, namely, the exponential of the Chern-Simons
form, \be \Psi_{\rm K}[A]= \exp(S_{\rm cs}) \ee where $S_{\rm
cs}=k\int_\Sigma (\frac{1}{2}A\wedge \d A+\frac{1}{3}A\wedge
A\wedge A)$. This state is known in the (loop) quantum gravity
community as the {\it Kodama} state, since it was first defined by
Kodama in the context of mini-superspace cosmological models
\cite{Kodama} (although it was first introduced by Jackiw in the
context of Yang-Mills theory \cite{jackiw}). This state has been
extensively studied by Smolin, who has argued that, in the case of
$A$ a self-dual connection, the Kodama state represents a vacuum
state centered around classical de Sitter space \cite{Smolin}. In
a recent paper Witten has listed some of the features that make
this state unsuitable as a realistic vacuum state in pure
Yang-Mills theory \cite{Witten}. In particular it is known that
negative helicity states have negative energies (and norms). For
an Abelian gauge theory, Witten has also constructed these states.
More recently, Smolin and Freidel have argued that in gravity, the
lessons from Yang-Mills theory do not necessarily apply: one has
to consider the so called ``reality conditions" \cite{Freidel}. In
gravity with self-dual variables, this means choosing a physical
inner product that renders the real observables Hermitian, and
might solve the problem of non-normalizability of the state
\cite{Freidel}.

On the other hand, there has been some interest in the quantum
theory of the Maxwell field, for self-dual connections, both in a
holomorphic representation \cite{Ashtekar:vz} and in the
corresponding loop representation \cite{ac:photon}. In this later
case it was found that the inner product for loop excitations was
given by the Gauss linking number (of suitably smeared loops). The
Linking number was also found to be prominent in the basic
commutation relations (CCR) for (real) electric and magnetic
fluxes \cite{aa:ac}.

The purpose of this note is twofold. On the one hand, we would
like to make precise the construction outlined in \cite{Witten}
for Abelian fields, and to point to the interesting role that
self-duality plays in the process. To be precise, we shall
construct the Schr\"odinger representation for the CCR
corresponding to a self-dual decomposition of the classical phase
space, and find that the canonical vacuum state is given by the
Kodama wave function. The decomposition leading to this state is
metric-background independent and in a sense explains why the
resulting quantum theory, with the Chern-Simons form as the vacuum
state, is topological. This leads us to the second goal of this
note, namely to make contact with the results found previously for
self-dual fields \cite{Ashtekar:vz,ac:photon}, and for the
commutation relations \cite{aa:ac}, where simple link invariants
were found to play key roles in the quantum theory. The interplay
between self-duality, Chern-Simons and diffeomorphism invariance
exhibited in this note allows us to put those previous results in
a coherent picture.

The structure of this paper is as follows. In Sec~\ref{sec:2} we
consider the self-dual decomposition of the phase space that leads
to the corresponding (Fock) quantization. In Sec.~\ref{sec:3} we
construct the corresponding Schr\"odinger representation, finding
that the vacuum state is the Kodama wave function. We conclude
this note in Sec.~\ref{sec:4} with a discussion.

Throughout this paper, we shall consider $c=1$, but will keep $e$
and $\hbar$ dimension-full.

\section{Self-Dual Quantization}
\label{sec:2}

Let us consider a $U(1)$ gauge theory, that can be, for
simplicity, thought  to be the Maxwell theory. In the canonical
approach the phase space $\Gamma$ is described by the magnetic
potential $A_a$ and the electric field $E^a$, each point of
$\Gamma$ defined by a pair $(A,E)$. For convenience, we introduce
the electric vector potential $V_a$ such that
$E^{a}=\epsilon^{abc}\nabla_{b}V_{c}$, and
$B^{a}=\frac{1}{\sqrt{h}}\tilde{\eta}^{abc}\underline{F_{bc}}$
 is the magnetic field on a hypersurface $\Sigma$. Recall that the
 basic Poisson Brackets are given by
 \be
 \{ A_a(x),E^b(y)\}=4\pi\delta^b_a\delta^3(x,y)
 \ee
We are using Gauss system of units for which $A_a$ carries the
dimensions of $\sqrt{M/L}$ and $E^a$ has dimensions of
$\sqrt{M/L^3}$. It is sometimes convenient to introduce smeared
observables for the basic variables, \be
A[f]:=\int_{\Sigma}{\rm{d}}^{3}\!x \, \sqrt{h}\,A_{a} f^{a} \quad
, \quad E[g]:=\int_{\Sigma}{\rm{d}}^{3}\!x \,\sqrt{h}\, E^{a}g_{a}
\ee

 For
these basic observables, the corresponding Canonical Commutation
Relations (CCR) read, \be [\hat{A}[f],\hat{E}[g]]=4\pi i \hbar
\int_{\Sigma}{\rm{d}}^{3}\!x \,\sqrt{h}\, f^{a}g_{a}\,
\hat{{\rm{I}}} \ee

It is known that in field theory on any curved background
(including Minkowski), there is an infinite freedom in the
representation of the CCR, or as is normally known, the choice of
vacuum state \cite{wald2}. In Minkowski space-time, Poincar\'e
invariance selects the standard vacuum, but in the absence of
symmetries the choice is not so straightforward.

In this note we will consider a particular representation that is
very natural from the geometric viewpoint, even when it is not
physically acceptable. This is the representation induced by the
Hodge-duality operator $\star$. Recall that the $\star$-operator
acts on the Faraday tensor $F_{ab}$ as
${}^{\star}F_{ab}=\frac{1}{2}\,\epsilon_{abcd}F^{cd}$. The
relevant feature of the duality operator is that when squared it
gives minus the identity: ${}^\star({}^\star F_{ab})=-F_{ab}$.
Therefore, the Hodge-$\star$ induces a complex structure $J$ on
the phase space \cite{ac:fock}. In terms of the canonical data,
the complex structure takes the form,
\begin{eqnarray}\
J\cdot\left( \begin{array}{c} A_{a} \\ E^{b} \end{array} \right)=
\left( \begin{array}{c} -V_{a} \\ B^{b} \end{array} \right)
\end{eqnarray}
One can also write it as the (matrix) operator,
\begin{eqnarray}\
J= \left( \begin{array}{cc} 0 & -(\nabla \times)^{-1} \\
\nabla \times & 0 \end{array} \right)
\end{eqnarray}
It is important to note that this self-dual decomposition of the
phase space is completely background independent (or as sometimes
is depicted, diffeomorphism invariant). This means that, given the
manifold $\Sigma$ and initial data $(A,E)$ on $\Sigma$ , the
complex structure $J$ provides a canonical decomposition,
independently of the metric $g_{ab}$ on the space-time $M$ and of
the particular embedding of $\Sigma$ into $M$. In the standard
Fock representation, the one particle Hilbert space ${\cal H}$ is
built by completing the phase space $(A,E)$ with respect to the
Hermitian inner product given by, \be
\langle\cdot,\cdot\rangle={{1}\over{2\hbar}}\mu(\cdot,\cdot)-
i{{1}\over{2\hbar}}\w(\cdot,\cdot)\, .\label{innp}
 \ee
 where the real inner product $\mu (\, \cdot \, , \, \cdot \,)=\w(J\,
\cdot \, ,  \, \cdot \,)$ \cite{wald2}. Here $\w(\, \cdot \, ,\,
\cdot \,)$ is the naturally defined symplectic two-form
\cite{covariant,ac:fock}, that for canonical data $(A,E)$ reads,
$$ \Omega((A,E),(A',E'))=\int_{\Sigma}\sqrt{h}\,{\rm{d}}^{3}\!x \,
(E^{a}A'_{a}-E'^{a}A_{a})\, .$$

\section{Kodama State}
\label{sec:3}

The corresponding Schr\"odinger representation can be easily
obtained by a direct generalization of the result of
Ref.\cite{nosotros} to the Maxwell field. Let us now summarize the
main steps in the construction. The first step is to consider the
generators of the Weyl algebra, \be
 \hat{W}[l]:=\exp\biggl[\frac{i}{\hbar}\hat{\Omega}(l,\, \cdot\,)\biggr]
 \ee
where \be \Omega(l, \,
\cdot\,)=\Omega((-g_a,f^b),(A_a,E^b))=A[f]+E[g] \ee is a general
linear observable labelled by the pair $(-g,f)$.

The strategy in order to construct the Schr\"odinger
representation is to employ a GNS construction for which an
algebraic state $\omega$ is defined as, \be
\omega(\hat{W}[l])=\exp\left[-\frac{\pi}{\hbar}\mu(l,l)\right] \ee
In the resulting quantum theory, the quantity $\omega(\hat{W}[l])$
corresponds to the vacuum expectation value of the Weyl generators
$\langle\hat{W}[l]\rangle_0$ . For details of the construction see
\cite{nosotros}, and for a summary and some discussion see
\cite{ccq:cqg}. In order to arrive at the quantum  measure, one
considers the Weyl generators corresponding to configuration
observables, in such a way that we have, \be \exp \biggl[  -
\frac{\pi}{\hbar} \mu((0,f^{a}),(0,f^{a}))\biggr] =
\int_{\bar{{\cal{C}}}}{\rm{d}}\mu \, \exp \biggl[  -
\frac{i}{\hbar} A[f]\biggr] \ee What is to say, that the Fourier
transform $\chi[f]$ of the measure $\d\mu$ given by, \be
\chi_{\mu}[f]=\int_{\bar{{\cal{C}}}}{\rm{d}}\mu \, \exp \biggl[ -
\frac{i}{\hbar} A[f]\biggr]\, . \ee
 takes the form
 \be
\chi_{\mu}[f]=\exp\left[-
 \frac{\pi}{\hbar}\int_{\Sigma}{\rm{d}}^{3}\!x\,\sqrt{h}
\,f^{a}(\epsilon^{abc}\nabla_{b})^{-1}f^{c}\right] \ee Now, the
general theory of measures \cite{draft} assures us that the
corresponding measure takes the form, $$ \mbox{``}{\rm{d}}\mu =
\exp \biggl[-\frac{1}{4\pi\hbar}\int_{\Sigma}{\rm{d}}^{3}\!x \,
\sqrt{h}\,
 \epsilon^{abc}A_{a}\,\nabla_{b}A_{c}
\biggr]\, {\cal{D}}A \mbox{''}, $$ But since
$\int_{\Sigma}{\rm{d}}^{3}\!x \,\sqrt{h}\,
\epsilon^{abc}A_{a}\nabla_{b}A_{c}=\frac{1}{2}\int_{\Sigma}A
\wedge \d A$, we have that,
 \be \mbox{``}{\rm{d}}\mu = \exp
\biggl[-\frac{1}{8\pi\hbar}\int_{\Sigma}A \wedge \d A \biggr]\,
{\cal{D}}A \mbox{''} \ee

If we consider a homogeneous measure, then the vacuum ``absorbs"
the measure and becomes non-trivial \cite{ccq:fock}. Thus we
arrive at the vacuum state given by, \be
 \Psi_{0}[A]=\exp \biggl[ -\frac{1}{16\pi\hbar} \int_{\Sigma}A
\wedge \d A \biggr]
 \ee
which is the so-called {\it Kodama state}. Note that we have been
working in the ``kinematical phase space", namely before imposing
Gauss' law. However, the resulting representation is gauge
invariant and naturally projects down to the reduced phase space.

Let us now explore this representation a bit. The basic
observables are represented as, \be
 \hat{A}[f]\cdot\Psi[A]=(A[f] \Psi)[A]\, ,
 \ee
 and
 \be
 \hat{E}[g]\cdot\Psi=-i4\pi\hbar\int_\Sigma{\rm{d}}^{3}\!x \,
 g_a\left(\frac{\delta}{\delta A_a}\,\Psi\right)
 \ee
Now, the vacuum state satisfies the following {\it self-duality}
property: \be (\hat{B}^a+i\hat{E}^a)\cdot \Psi_{0}[A]=
\left[B^a+i\left(-4i\pi\hbar\frac{\delta}{\delta
A_a}\right)\right]\Psi_{0}[A]= 0 \ee

This is the expected behavior of the Kodama state. Furthermore, it
is an eigenstate of the Hamiltonian with zero eigenvalue as
expected \cite{Witten}. It is important to note that this
self-dual property of the quantum vacuum state is satisfied for
the Kodama state for {\it real} U(1) connections (as opposed to
self-dual connections \cite{Ashtekar:vz}).

Let us now consider Wilson loops functions $h_\alpha[A]$, labelled
by loops $\alpha$, of the form \be
h_\alpha[A]=\exp\left(i\frac{e}{\hbar}\oint_\alpha A\right)
 \ee
An obvious question one can ask is  for the vacuum expectation
value of such operators,
 \be
\langle \hat{h}_\alpha\rangle_0= \int_{\bar{{\cal{C}}}}{\cal D}A\,
\exp\biggl( -\frac{1}{8\pi\hbar} \int_{\Sigma}A \wedge \d A
\biggr)\,\exp\left(i\frac{e}{\hbar}\oint_\alpha A\right) \ee
 This question has been extensively discussed in the literature (for a
pedagogical account see \cite{knots}), and it is agreed that,
after a framing, the quantity coincides with the self-linking of
$\alpha$. If we now consider the ``two loop function", $\langle
\hat{h}_\alpha\,\hat{h}_\beta\rangle_0$, one should expect that
this quantity should also be a knot invariant, related to the
Gauss Linking number of the loops $\alpha$ and $\beta$
\cite{knots}. We recover then the intriguing result that the inner
product between loop states is a knot invariant, when using
self-dual connections \cite{ac:photon}. That this quantum theory
incorporates the linking number can be seen from the fact that the
inner product between two 1-loop states
$\hat{A}[\alpha]|0\rangle:=\widehat{(\oint_\alpha A)}|0\rangle$
and $\hat{A}[\beta]|0\rangle:=\widehat{(\oint_\beta A)}|0\rangle$
is given by the $\mu$ inner product on ${\cal H}$ \cite{ccq:fock}.
This quantity then takes the form, \ba\label{2-loop} \langle \hat{
A}[\alpha]\,\hat{A}[\beta]\rangle_0 &=&
\frac{2\pi}{\hbar}\,\mu(F[\alpha],F[\beta])\\ &=&
\frac{2\pi}{\hbar}\,\int_\Sigma \d
^3\!x\,F^a[\alpha,x)(\epsilon^{abc}\nabla_{b})^{-1}F^{c}[\beta,x)\nonumber
 \ea
 where $F^a[\alpha,x)=\oint_\alpha\d
\alpha^a\delta^3(\alpha,x)$ is the form factor of the loop
$\alpha$, such that $\int_\Sigma\d^3\!x\,A_a(x)F^a[\alpha,x)=
\oint_\alpha A$.  In Ref. \cite{aa:ac}  it was shown that the
Right Hand Side of (\ref{2-loop}) is indeed given by the Gauss
linking number of $\alpha$ and $\beta$. Thus, we get:  \be \langle
\hat{A}[\alpha]\,\hat{A}[\beta]\rangle_0=\frac{2\pi}{\hbar}{\cal
GL}(\alpha,\beta)\,. \label{gl2} \ee
 One could also consider the so called
r-holonomies --- namely suitably smeared holonomies---, as the
relevant configuration observables \cite{mad1}. One would then
get, instead of (\ref{gl2}), properly generalized linking numbers
as in \cite{ac:photon}.

\section{Discussion}
\label{sec:4}

Let us summarize our construction. We started with a complex
structure on phase space defined by the Hodge-dual on the
space-time and, by means of a GNS construction, ended with the
vacuum state given by the Kodama State. From this perspective, it
is clear why this state has several problems. To begin with, the
duality complex structure $J$ is not compatible with the
symplectic structure $\Omega$, which means that the corresponding
one particle Hilbert space ${\cal H}$ in the Fock construction has
negative norm states. Furthermore, the operators that define the
Fourier transform of the measure do not satisfy the positivity
requirements: the measure is non-Gaussian, and non-normalizable.

Perhaps the most important question is whether these results on
Abelian gauge theories can teach us something about the
non-Abelian case, and in particular about gravity. Clearly we can
say nothing about the issue of how a physical inner product on the
space of solutions to the constraints might affect the
normalizability of the state. However, we can still learn
something from this simple model. What we have seen is that we do
not need to introduce self-dual variables and complicated reality
conditions in order to get a ``self dual" representation and
vacuum state \cite{Ashtekar:vz}; our variables were always real.
The self-duality in the resulting quantum theory came from the
quantum representation, in this case due to the choice of complex
structure on phase space. One might hope that this new perspective
on the problem may shed some light in the goal of constructing
physically relevant vacuum states and representations for gravity,
without the need to introduce self-dual complex connections. Note
also that even when we have considered a pure Abelian gauge field,
this construction can also be done for linearizations of
Yang-Mills \cite{Freidel} and gravity theories \cite{mad}.

Let us end with two remarks:

\begin{enumerate}

\item In retrospect, one might argue that the fact that one
recovers the results of \cite{ac:photon} is not entirely
unexpected. After all, the Schr\"odinger representation with a
vertical polarization for the Hodge-dual complex structure {\it
should} be equivalent to a holomorphic representation with the
self-dual polarization as in \cite{Ashtekar:vz} and
\cite{ac:photon}.

\item Nevertheless, it is interesting to see the non-trivial
interplay between self-duality and diffeomorphism invariance that
this model possesses. It has been known for some time that the
symplectic structure $\Omega(\, ,\,)$ that appears in the inner
product (\ref{innp}) is diffeomorphism invariant, a fact that is
reflected in the Commutation Relations of the theory for the right
kind of (electric and magnetic flux) variables \cite{aa:ac}. In
this note we have seen that when the other part of
Eq.~(\ref{innp}), namely the real inner product $\mu(\, ,\,)$, is
also diffeomorphism invariant, then we recover a fully topological
quantum theory, at least at the kinematical level.

\end{enumerate}

\begin{acknowledgments}
 This work was  supported in part by
DGAPA-UNAM grant No. IN112401 and by CONACyT grant J32754-E.

\end{acknowledgments}


\begin{thebibliography}{99}





\bibitem{Kodama}
H.~Kodama, ``Holomorphic Wave Function Of The Universe,'' Phys.\
Rev.\ D {\bf 42}, 2548 (1990).

\bibitem{jackiw} R. Jackiw, ``Topological Investigations in
Quantized Gauge Theories" in {\it Current Algebra and Anomalies},
Ed. S.B. Treiman {\it et al} (World Scientific, 1985).

\bibitem{Smolin}
For a recent review see: L.~Smolin, ``Quantum gravity with a
positive cosmological constant,'' arXiv:hep-th/0209079.

\bibitem{Witten}
E.~Witten, ``A note on the Chern-Simons and Kodama
wavefunctions,'' arXiv:gr-qc/0306083.

\bibitem{Freidel}
L.~Freidel and L.~Smolin, ``The linearization of the Kodama
state,'' arXiv:hep-th/0310224.

\bibitem{Ashtekar:vz}
A.~Ashtekar, C.~Rovelli and L.~Smolin, ``Selfduality And
Quantization,'' J.\ Geom.\ Phys.\  {\bf 8}, 7 (1992)
[arXiv:hep-th/9202079].

\bibitem{ac:photon}
A.~Ashtekar and A.~Corichi, ``Photon inner product and the
Gauss-linking number,'' Class.\ Quant.\ Grav.\  {\bf 14}, A43
(1997) [arXiv:gr-qc/9608017].

\bibitem{aa:ac}
A.~Ashtekar and A.~Corichi, ``Gauss linking number and
electro-magnetic uncertainty principle,'' Phys.\ Rev.\ D {\bf 56},
2073 (1997) [arXiv:hep-th/9701136].

\bibitem{wald2} R.M. Wald, {\em Quantum Field Theory in Curved Space-time
and Black Hole Thermodynamics} (Chicago University Press, Chicago,
1994).

\bibitem{ac:fock}  A. Corichi, Rev. Mex. Fis. {\bf 44}, 402 (1998).
[arXiv:physics/9804018].

\bibitem{covariant} C. Crnkovic and E. Witten, in {\it Three hundred
years of gravitation}, Cambridge U. Press (1987).


\bibitem{nosotros} A. Corichi, J. Cortez, H. Quevedo,
``Schr\"odinger representation for a scalar field on curved
spacetime,'' Phys.\ Rev.\ D {\bf 66}, 085025 (2002)
[arXiv:gr-qc/0207088].

\bibitem{ccq:cqg} A.~Corichi, J.~Cortez and H.~Quevedo,
``Note on canonical quantization and unitary equivalence in field
theory,'' Class.\ Quant.\ Grav.\  {\bf 20}, L83 (2003)
[arXiv:gr-qc/0212023].

\bibitem{draft}
Y.Yamasaki, {\it{Measures on infinite dimensional spaces}} (World
Scientific, 1985).

\bibitem{ccq:fock}
A.~Corichi, J.~Cortez and H.~Quevedo, ``On the relation between
Fock and Schr\"odinger representations for a  scalar field,''
arXiv:hep-th/0202070.


\bibitem{knots} L.H. Kauffman, {\it Knots and Physics} (World
Scientific, 1991)



\bibitem{mad1}
M.~Varadarajan, ``Fock representations from U(1) holonomy
algebras,'' Phys.\ Rev.\ D {\bf 61}, 104001 (2000)
[arXiv:gr-qc/0001050].

\bibitem{mad} M.~Varadarajan,
``Gravitons from a loop representation of linearised gravity,''
Phys.\ Rev.\ D {\bf 66}, 024017 (2002) [arXiv:gr-qc/0204067].






\end{thebibliography}
\end{document}